\documentstyle[prl,aps,epsf]{revtex}
\begin{document}
\twocolumn[\hsize\textwidth\columnwidth\hsize
     \csname @twocolumnfalse\endcsname
\draft

\title{The origin of the 90 degree magneto-optical Kerr rotation in CeSb}
\author{U. Pustogowa, W. H\"ubner, and K. H. Bennemann}
\address{Institute for Theoretical Physics, Freie Universit\"at 
Berlin, Arnimallee 14, D-14195 Berlin, Germany}
\date{\today}
\maketitle
\begin{abstract}
We calculate the {\em linear} magneto-optical Kerr rotation for CeSb 
in the near-infrared spectral range. 
Using an exact formula for large Kerr rotation angles and a simplified 
electronic structure of CeSb we find at $\hbar \omega = $ 0.46~eV 
a Kerr rotation of 90$^{\circ}$ which then for decreasing $\omega $ jumps to 
-90$^{\circ }$ as recently observed. 
We identify the general origin of possible 180$^{\circ }$ polarization 
rotations as resulting from mainly nonmagnetic optical properties, 
in particular from the ratio of the dominant interband resonance frequency to 
the plasma frequency. 
The dependence of the Kerr rotation on moments and magnetization is discussed. 
\end{abstract}
%
%

\vskip1pc]
\newpage
\narrowtext

The search for large Kerr rotations in magneto-optics has been a longstanding 
subject of both basic and application-oriented research. 
The prediction and observation of giant Kerr 
rotations~\cite{pusto,Kirsche,prlKoop,Hue95} in the nonlinear magneto-optical 
Kerr effect (NOLIMOKE) on multilayer sandwiches and thin magnetic films have 
successfully demonstrated the enhanced sensitivity of nonlinear optics to 
magnetism, in particular on low-dimensional systems due to the reduction of 
symmetry. 
Although the detailed values of the enhanced Kerr rotation in NOLIMOKE depend 
on the electronic structure of the investigated system, the spin-orbit 
coupling strength, 
the magnetization direction in the sample and the polarization of the 
incoming light,  
the 90$^{\circ}$ nonlinear Kerr rotation in the longitudinal 
configuration and steep angle of incidence, is essentially independent of the 
frequency and results from symmetry considerations~\cite{Hue95}.  

Prior and also parallel to NOLIMOKE, the search for enhanced Kerr rotations 
has been pursued in the conventional linear magneto-optic Kerr effect 
(MOKE). 
Since the MOKE rotation of transition metals like Fe or Ni is typically in the 
range of only some 0.1$^{\circ}$ for optical frequencies one had to resort to 
particular magnetic rare-earth or uranium alloys with large spin-orbit 
coupling constants, large magnetic moments, and with particular optical 
resonances in only one spin type to find relatively large Kerr rotations 
at low temperatures ($\approx $~1~K) and at lower frequencies. 
Furthermore, the application of large external magnetic fields 
($\approx $~5~T) was necessary. 
In this way, a record-high MOKE rotation of 14$^{\circ}$ has been observed 
for CeSb in 1986 by Schoenes {\em et al.}~\cite{Reim86}. 
Recently, however, the same group (Pittini {\em et al.}~\cite{moris}) 
observed the largest possible rotation of 90$^{\circ}$ and an abrupt jump of 
this rotation to -90$^{\circ }$ in CeSb by reducing the frequency from 
0.55 eV to 0.46 eV. 

Up to now, two groups performed first principle calculations of Kerr spectra 
of CeSb~\cite{Opp,Liecht} both based on LDA+U electronic structure 
calculations. Both emphasize as most important for near-infrared optics 
the interband transitions in hybridized Sb $p$ bands and get similar 
diagonal and off-diagonal conductivities. 
Yaresko {\em et al.}~\cite{Opp} used an exact expression for large Kerr 
rotation angles but did not reproduce the $\pm 90^{\circ }$ rotation jump. 
Liechtenstein {\em et al.}~\cite{Liecht} did not calculate the Kerr rotation. 


Thus, our aim is to explain the $\pm 90^{\circ }$ Kerr rotation in terms of 
the susceptibility tensor $\chi_{ij}$ and, in particular, 
to discuss the influence of interband and intraband parts of the electrical 
susceptibility using typical electronic properties of CeSb. 
We argue that this large Kerr rotation occurs for a frequency, which is set by
both the plasma frequency and special optical transitions allowed only for one 
spin type. 
Thus, the Kerr rotation is not directly related to the magnetic properties. 
Our explanation relies largely on an improved evaluation of the Kerr 
rotation for a model band structure which does not assume any small 
parameters in the off-diagonal components of the reflected electric field. 
We use the result of LDA+U calculations, in order to understand 
the effect of the Ce $f$ electrons spin-polarizing and splitting up the Sb 
$p$ bands via hybridization. 
Our simplified theory outlined in the following attempts to describe 
the physical origin of the large linear Kerr effect. 

The relation of MOKE signals (Kerr rotation or intensity measurements) to 
magnetic properties of the material is of special interest, 
since MOKE is used for the determination of the direction and relative 
strength of the magnetization in saturated ferromagnetic materials. 
By analyzing the reflectivity, the magnetic dichroism, and the 
electrical-susceptibility-tensor elements, we show that this special
90$^{\circ }$ Kerr rotation is based on the behavior of the diagonal 
'nonmagnetic' susceptibility $\chi_{xx}(\omega )$ and thus is not  
related to the magnitude of the magnetization in the sample. 

To determine large Kerr rotation angles we have to use general 
expressions for the Kerr rotation $\varphi $ and ellipticity in contrast to 
the usually taken linearized formulae which are valid if 
$\tan \varphi \approx \varphi $ holds. 
Following general ellipsometry arguments the complex polar Kerr angle 
$\kappa $ is defined as
	\begin{equation}
	\kappa \; = \; \frac {E_{y}} {E_{x}} \; = \; 
	\frac {\tan \varphi + i \tan \varepsilon }
		{1-i \tan \varphi \tan \varepsilon } \; ,
	\label{glw1}
	\end{equation}
where, $\varphi $ is the Kerr rotation angle and  
$\varepsilon $ is the ellipticity. 
$E_{y}$ and $E_{x}$ are components of the reflected electric field. 
Solving Eq.~(\ref{glw1}) we find for the Kerr angle   
	\begin{equation}
	\varphi \; = \; \arctan \left( - \frac {1- |\kappa |^{2} } 
	{2 \mbox{Re}(\kappa ) } \pm 
	\sqrt {\frac {(1- |\kappa |^{2})^{2} } 
	{4 [\mbox{Re}(\kappa )]^{2} } + 1} \right) \; .
	\label{loesung}
	\end{equation} 
The analysis of Eq.~(\ref{loesung}) yields that 180$^{\circ }$-changes in 
$\varphi $ may occur when Re($\kappa $) is zero or infinity. 
The behavior of $\kappa $ results from 
the well known expression for the linear polar Kerr rotation 
	\begin{equation}
	\kappa \;=\;-\frac{\chi _{xy}^{(1)}(\omega )}
	{\chi _{xx}^{(1)}(\omega )}
	\frac{1}{\sqrt {1 + \chi _{xx}^{(1)}(\omega )}} \; .
	\label{lin}
	\end{equation}
Here, $\chi _{xx}^{(1)}$ and $\chi _{xy}^{(1)}$ denote the elements of the 
linear susceptibility tensor~\cite{einheit}.
The complex value $\kappa $ describes the tangent of the angle, 
$\kappa = \tan \Phi_{K}$, with $\Phi_{K}=\varphi + i \varepsilon \,$. 
Note, for small Kerr rotations Eq.~(\ref{loesung}) reduces to 
$\varphi $ = Re($\kappa $). 
Eq.~(\ref{loesung}) can be rewritten as 
	\begin{equation} 
	\varphi \; = \; \frac {1} {2} \arctan \left( 
	\frac {2 \mbox{Re}(\kappa )} {1-|\kappa |^{2}}\right) + 
	\varphi_{0} \; ,
	\label{glphi0}
	\end{equation}
with $\varphi_{0} = 0 $ for $ |\kappa |^{2} \le 1 $, 
$\varphi_{0} = 90^{\circ} $ for $ |\kappa |^{2} > 1,$  Re$(\kappa ) \ge 0 $, 
and $\varphi_{0} = -90^{\circ} $ for $ |\kappa |^{2} > 1,$ Re$(\kappa ) < 0$.
The ellipticity angle $\varepsilon$ is given by 
	\begin{equation}
	\varepsilon \; = \; \; \frac {1} {2} \arcsin \left( 
	\frac {2 \mbox{Im}(\kappa )} {1+|\kappa |^{2} } \right) \; .
	\label{epsi}
	\end{equation}
For details see Groot Koerkamp~\cite{Koerdipl,foot}. 

Thus, the behavior of $\varphi $ for Re$(\kappa ) \rightarrow 0 $ depends on 
the value of $|\kappa |^{2}$. 
For $|\kappa |^{2} \le 1$, the case of small Kerr rotations, 
together with Re$(\kappa )$ also $\varphi $ goes to zero. 
In the other case, for $|\kappa |^{2} > 1$ , changing the sign of Re$(\kappa )$ 
yields two different values of $\varphi$, including a jump from 
$90^{\circ}$ to $-90^{\circ}$, since the Kerr rotation is defined only 
modulo $\pi $.  

So far, Kerr effect measurements realized only the first case 
$|\kappa |^{2} \le 1$ and no $\pm 90^{\circ}$ rotation angle was observed. 
Thus, we define two conditions for the occurence of a $\pm 90^{\circ }$ 
rotation angle $\varphi $
	\begin{equation}
	\mbox{Re}(\kappa )=0 \qquad \mbox{and} \qquad 
	|\mbox{Im}(\kappa )| > 1
	\label{condition}
	\end{equation}
Note, this was not explicitely discussed previously~\cite{rasing}. 
Here, these conditions are formulated for the calculation-related values 
$\kappa $ and $\chi $. 
By means of the reflected perpendicularly polarized electric fields $E_{y}$ and 
$E_{x}$, Eq.~(~\ref{condition}) reproduces a complex phase of $90^{\circ }$ 
(see Re$(\kappa ) =0$) and $E_{y0} > E_{x0}$ leading to $|\kappa | >1$. 

In the following we describe a model for the electronic structure of CeSb. 
Thus, we analyze Eqs.~(\ref{glphi0}) and (\ref{epsi}) for realistic 
values of $\chi_{xx}$ and $\chi_{xy}$ in contrast to the discussions 
in refs.~\cite{prlKoop,Hue95} based on symmetry considerations. 

For describing now the infrared Kerr-spectrum of CeSb as measured by Pittini 
{\em et al.}~\cite{moris} we use {\em only one} feature of the electronic 
structure of CeSb close to the Fermi level,  
namely fairly flat and nearly parallel bands above and below the Fermi level 
$\varepsilon_{F}$ in the $\Gamma $ -- Z direction~\cite{Liecht}.
The $p$ states of Sb are split by approximately 0.6 eV. 
Furthermore, via hybridization with the $f$-electron minority-spin subband 
of CeSb (lying $\sim 3$ eV above $\varepsilon_{F}$) the Sb $p$ states 
have an induced spin polarization. 
Thus, there are favored transitions of one spin sort between the $p$ 
states of very high weight~\cite{dipol}. 
In this case, the interband parts of the diagonal and off-diagonal 
susceptibilities can be written as sums over transitions for one spin 
band only. 
Thus, 
	\begin{equation} 
	\chi_{xx} = \chi_{xx,intra} \; + \; 
	\sum_{l,l^{\prime}} L_{l,l^{\prime}} \; , 
	\; \; \; \; \; 
	\chi_{xy} = - i 
	\frac {\lambda_{s.o.} } {\hbar \omega} 
	\sum_{l,l^{\prime}} L_{l,l^{\prime}} \; ,
	\end{equation}
where $\chi_{xx, intra}$ denotes the intraband contribution important only 
in the diagonal elements of $\chi $,   
$L_{l,l^{\prime}}$ denotes the 
response from transitions between the bands $l$ and $l^{\prime }$ and 
$\lambda_{s.o.}$ is the spin-orbit coupling constant. 
Note, the factor $\lambda_{s.o.}/\hbar \omega $ results from including 
spin-orbit coupling to lowest order in the wave functions. 
However, this will not directly affect $\varphi $. 
The Kerr rotation is then calculated by using 
	\begin{equation}
	\kappa \;=\; i \, \frac{\lambda_{s.o.}}
	{\hbar \omega } \; 
	\frac{1}{ \left( 1+ \frac {\chi_{xx,intra}} {L} \right) \; 
	\sqrt {1 + \chi _{xx}^{(1)}(\omega )}} \; .
	\label{kapl}
	\end{equation}
For simplicity, we approximate the transitions between the (Sb) $p$ bands 
by a single atomic (nondispersive) Lorentzian
	\begin{equation}
	\sum_{l,l^{\prime}} L_{l,l^{\prime}} \; = \; L \; := \; 
	\frac {C} {E_{f} - E_{i} - \hbar \omega + i \hbar \alpha } 
	\; ,
	\end{equation} 
with the band positions $E_{f}$ of the unoccupied final state and $E_{i}$ of 
the occupied initial state, the damping factor $\alpha $, and the 
coefficient $C = (e^{2}M^{2}/V) $. 
In metals the symmetry of the transition-matrix elements $M$ plays a less 
important role and we use for simplicity the value $C=1$. 
Note, optical transitions between {\em minority}-spin-electron states 
dominate. 
Furthermore, we describe intraband effects by a conventional Drude term 
	\begin{equation}
	\chi _{xx, intra} = - \frac {\omega_{pl}^{2}} 
	{ \omega (\omega + i\tau ) } \; ,
	\end{equation}
with the plasma frequency $\omega _{pl}$ and the damping $\tau $. 
Using these expressions one can then determine $\varphi $ with the help of 
Eq.~(\ref{glphi0}). 

For a further analysis and for comparison with experiment we also calculate 
the optical reflectivity R using 
	\begin{equation}
	R \; = \; \left| \frac {N-1} {N+1} \right| ^{2} \; ,
	\end{equation}
with the refraction index $N^{2} = \epsilon_{0} + i\epsilon_{1} = 
1 + \chi_{xx} + i\chi_{xy}$.

We now present results for the Kerr rotation, the reflectivity, and 
in particular the dependence of the large Kerr angle on the plasma 
frequency $\omega_{pl}$ and interband splitting $\Delta E = E_{f} - E_{i}$ 
referring to the dominant optical transition. 

In Fig. 1 the frequency dependence of the linear polar Kerr angle $\varphi $ 
and the ellipticity $\varepsilon $ are shown. 
Here $\varphi $ and $\varepsilon $ are calculated using Eqs.~(\ref{glphi0}), 
(\ref{epsi}), and (\ref{kapl}). 
Parameters are choosen such as to obtain the change of $\varphi $ 
from -90$^{\circ }$ to +90$^{\circ }$ at precisely 0.46~eV as measured by 
Pittini {\em et al.}~\cite{moris}. 
Thus, we use for the interband splitting $\Delta E$ = 0.67~eV 
in agreement with the bandstructure result of Liechtenstein 
{\em et al.}~\cite{Liecht} and for the damping $\hbar \alpha $ = 0.1~eV as 
usually taken for Kerr angle calculations. 
In the Drude 
\vspace*{-8.5cm}
\begin{figure}
\epsfxsize=10.2cm
\centerline{\quad \epsfbox{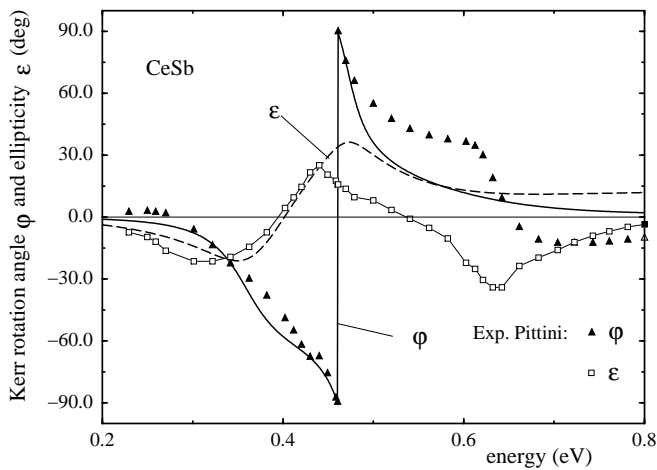}}
\vskip 1.0cm
\caption[]{Calculated frequency dependence of the linear magneto-optical 
Kerr-angle $\varphi$ (solid curve) and the ellipticity angle $\varepsilon$ 
(dashed curve) of CeSb calculated with a Lorentz-type susceptibility.
For comparison also the experimental results of Pittini {\em et al.} [6] for 
$\varphi (\hbar \omega)$ and $\varepsilon (\hbar \omega )$ are shown.}
\end{figure}
\bigskip
\noindent
term the parameters $\hbar \omega_{pl}=$~0.93~eV 
and $\hbar \tau = 0.95 \times 10^{-4}$~eV are used. 
The plasma frequency is comparable with the calculated value of 1.1~eV 
(see Ref.~\cite{Opp}). 
The value for $\tau $ is taken from Kwon {\em et al.}~\cite{KwonS}, but also 
values up to 10$^{-1}$~eV do not change the spectra. 
Note, the jump at 0.46~eV is really a jump from an angle -90$^{\circ }$ to 
an angle +90$^{\circ }$ without intermediate values   
in contrast to the measurement with an external field of 3 T~\cite{moris} 
(originated by the different magnetic phase of CeSb at 3~T and 5~T) and 
in contrast to {\em ab initio} electronic structure based 
calculations~\cite{Opp}. 
The ellipticity angle changes sign at 0.41~eV going from a minimum of 
-20$^{\circ }$ to the maximum of 35$^{\circ }$ . 
For comparison the experimental values by Pittini {\em et al.}~\cite{moris} 
are shown as dots. 

In Fig. 2 we show furthermore 
the frequency dependence of the optical reflectivity $R(\hbar \omega )$ 
in the energy range from 0 to 0.8~eV. 
For the frequency dependence of the Kerr angle the most important range of this
curve is the deep minimum at 
0.46~eV, at the same energy where the jump in 
$\varphi $ occurs. 
The discrepancy with respect to experimental results is presumably 
due to our simplified electronic structure of CeSb 
since we neglect interband transitions at other frequencies.  

\vspace*{-8.5cm}
\begin{figure}
\epsfxsize=10.2cm
\centerline{\quad \epsfbox{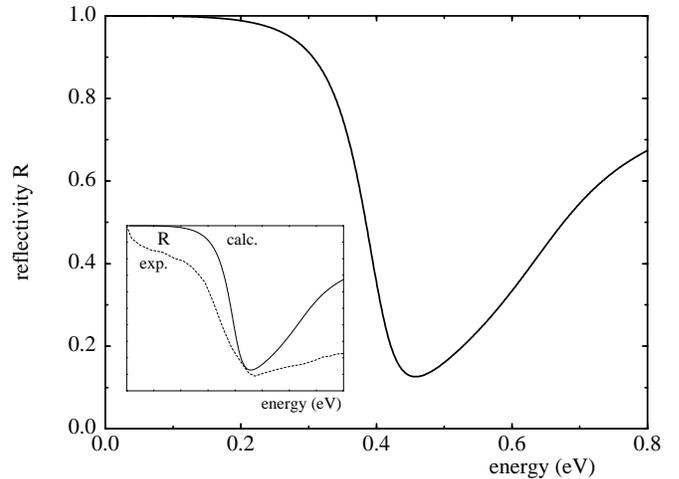}}
\vskip  1.0cm
\caption{Optical reflectivity $R$ of CeSb calculated from the diagonal 
susceptibility $\chi_{xx}(\omega)$. Experimental values are also shown.}
\end{figure}
\bigskip

Figs. 3 and 4 show the dependence of $\varphi (\hbar \omega )$ and, in 
particular, of the jump position and occurrence on $\Delta E$ and the plasma 
frequency $\omega_{pl}$. 
Fig. 3 shows Kerr-angle spectra $\varphi(\hbar \omega)$ with plasma 
frequencies varying from 0.2~eV to 1.3~eV. 
Note, for plasma frequencies higher than a threshold value depending on 
$\Delta E$ the 180$^{\circ }$-jump vanishes, 
whereas for low plasma frequencies the jump amplitude is stable, while only 
the position of the jump moves to lower energies. 
The variation of the interband transition $\Delta E$ yields the 
opposite behavior, see Fig. 4. 
Here, $\Delta E$ changes from 0.2 eV to 1.6 eV. 
The jump in $\varphi $ vanishes for low $\Delta E$ and moves for 
large values of $\Delta E$ to higher energies. 
Thus, we find that the ratio of the plasma frequency $\omega_{pl}$ to 
$\Delta E$, which characterizes the interband transitions, 
is essential for the occurrence of a 180$^{\circ }$ jump in the Kerr angle. 

The position of the 180$^{\circ }$ abrupt change in the Kerr angle $\varphi $ 
is determined by a zero of the real part of $\chi_{xx} $, in particular  
$\left( \chi_{xx,intra} + L \right) = 0$ . 
This abrupt Kerr-angle change may occur. 
Note, this equation above reflects the ratio $\frac {\chi_{xx}} {\chi_{xy}}$. 
We used two approximations, namely 
(i) $\chi_{xy} \sim \frac {\lambda_{s.o.}}{\hbar \omega }$ and 
(ii) assume interband transitions only in 
one spin subband. 
Within our band-structure model approximately for 
$\omega_{pl} / \Delta E \le 1.5 $ the abrupt Kerr angle change disappears. 
Moreover, there is no lower boundary for the occurrence of this change of the 
Kerr angle as a function of the ratio 
$\omega_{pl} / \Delta E$~\cite{Bemerkung}. 

For all parameter sets we analyzed the realization of the conditions in 
Eq.(~\ref{condition}). 
The first one, Re$(\kappa )=0$ was fulfilled in every case 
(at different frequencies) indicating a resonance-like behavior in $\varphi $. 
The additional analysis 
\vspace*{-8.0cm} 
\begin{figure}[h]
\epsfxsize=10.2cm
\centerline{\quad \epsfbox{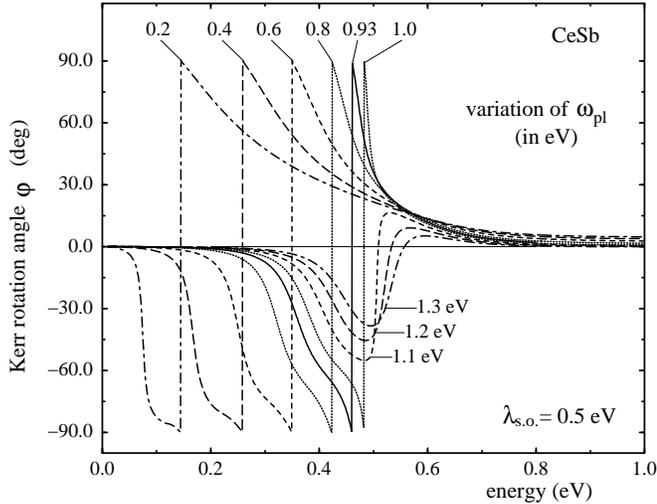}}
\vskip 1cm
\caption{Dependence of the Kerr angle spectra $\varphi (\hbar \omega)$ 
of CeSb on the value of the plasma frequency $\omega _{pl}$. 
For every curve the value of $\omega_{pl}$ (in eV) is shown.}
\end{figure}
\bigskip
\noindent
of the value of $|\kappa |$ or, since Re$(\kappa )=0$,
for Im$(\kappa )$, enables us to distinguish between the already discussed 
resonances by Feil and Haas~\cite{Feil} for $|$Im$(\kappa )| \le 1$ 
and the 'degenerate' resonances with the $\pm 90^{\circ }$ jump in $\varphi $  
for $|$Im$(\kappa )| > 1$.

\vspace*{-8.5cm}
\begin{figure}
\epsfxsize=10.2cm
\centerline{\quad \epsfbox{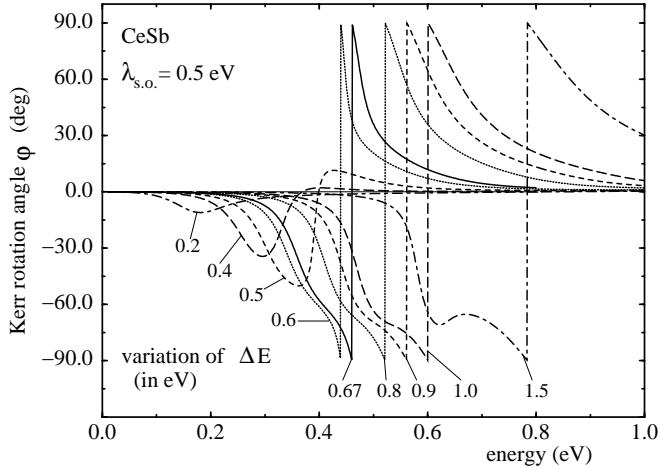}}
\vskip 1.0cm
\caption{Dependence of the Kerr angle spectra $\varphi (\hbar \omega)$ 
on the interband transition energy $\Delta E$. 
For every curve the corresponding value of $\Delta E$ (in eV) is shown .}
\end{figure}
\bigskip

Note, the $\pm 90^{\circ }$ change in the Kerr rotation can be described also 
by a continuous rotation from 0$^{\circ }$ to 180$^{\circ }$, which are 
identical polarization angles. 
Then, the condition $|$Im$(\kappa )| = 1$ marks a distinct boundary between 
the two cases: (i) $|$Im$(\kappa )| \le 1$, then the polarization shows a 
resonance-like behavior at Re$(\kappa )=0$ and returns back to the incident 
orientation direction 
and (ii) $|$Im$(\kappa )| > 1$, then a continuous rotation from $0^{\circ }$ 
to $180^{\circ }$ is allowed. 
The jump occurs only as a result of implementing the usual boundary conditions: 
$\varphi \rightarrow 0$ rotation angles outside the frequency region where 
Re$(\kappa )=0$. 
 
In the energy range below $\approx $ 0.5 eV our results for the Kerr angle 
$\varphi (\hbar \omega)$ and Kerr ellipticity $\varepsilon (\hbar \omega )$ 
reproduce the experimental data very well. 
The deviation for larger energies results from the neglect of further 
interband transitions. 
Note, we have included only one interband transition with $\Delta E$ in our 
model. 
For the same reason, as already mentioned, our calculation yields a too large 
reflectivity at 0.8 eV. 
The inclusion of more interband transitions, especially of transitions of the 
other spin, would decrease the reflectivity in this energy range and yield 
results for the Kerr rotation, ellipticity and reflectivity 
in better agreement with experiment for larger energies. 

Nevertheless, for the explanation of the 180$^{\circ}$ Kerr angle jump it is
not primarily important to know the exact electronic structure, but to fulfil
the condition Eq.(~\ref{condition}). We do this by combining the important features of the 
electronic structure with the plasma frequency. 

Historically, large linear Kerr rotations have been expected in materials 
with large spin-orbit coupling as in CeSb. 
In our model the spin-orbit coupling strength does of course influence the 
Kerr angle, since a large spin-orbit coupling is necessary to realize 
Im$(\kappa ) > 1$, but does not influence the position determined 
by Re$(\kappa) =0$ at which the Kerr rotation changes abruptly. 
In addition, in the case of CeSb the interband splitting $\Delta E$ is 
affected by the spin-orbit coupling. 
Note, for the Kerr effect in transition metals the spin-orbit interaction does 
not appreciably influence the electronic bandstructure, but of course has to be 
included in the wave functions. 

Note, the origin of the 180$^{\circ }$ Kerr rotation was discussed 
qualitatively in ref.~\cite{moris}. 
Dominant dipole transitions of only one spin type and low values of the 
optical constants were assumed. 
The first point we use directly in our model of the electronic structure of 
CeSb, whereas concerning the optical constants we formulate the exact 
conditions Eq.~(6) and relate them quantitatively to $\chi_{ij}$ and the 
electronic structure. 
In contrast to ref.~\cite{moris} we show the importance of the 
intraband part for the 180$^{\circ }$ rotation, especially the 
quantitative relation of intra- to interband transitions at the 'jump' 
frequency required for the realization of the jump conditions, 
as illustrated in Figs. 3 and 4. 
However, note the formalism used in ref.~\cite{moris} for the discussion of 
large Kerr rotation angles is based on the same electrodynamical expressions 
as in our derivation. 

In view of our model, related materials with pronounced interband 
transitions due to flat $f$ and $d$ bands could exhibit similar behavior. 
However, due to the threshold character of the $\varphi $ enhancement 
and the Kerr-angle jump, slight changes of the parameters and dominant optical 
transitions might already suppress the giant Kerr effect.
This may explain why for example CeBi with similar $\lambda_{s.o.} $ and 
magnetic moments exhibits only a Kerr angle of about 
-9$^{\circ }$ ~\cite{Pitti2}.
The different shapes of $\sigma_{xy} (\omega )$ for CeSb and CeBi might suggest 
already differences in the dominant dipole transitions. 
In addition, as discussed before, these dominant interband transitions in CeSb 
are explained by hybridized $p$ bands of Sb. 

Most interesting is the relation of the discussed Kerr-rotation jump 
to the magnetic properties of the material. 
In view of this we calculate the magnetic contrast $d$ for MOKE, 
	\begin{equation}
	d \; = \; \frac {I(+M)-I(-M)} {I(+M)+I(-M)} \; ,
	\end{equation}
where $I(\pm M)$ are the intensities for reversed magnetization directions. 
Using for the intensities of the reflected light 
$I=|E_{refl}|^{2} = |\chi  E_{incident}|^{2} $ we find 
$I(\pm M) \sim |\chi_{xx} \pm i \chi_{xy}|^{2}$ and for the magnetic contrast
	\begin{equation}
	d \; = \; 2 
	\frac { Im(\chi_{xx}) Re(\chi_{xy}) - Re(\chi_{xx}) Im(\chi_{xy}) } 
	{|\chi_{xx}|^{2} + |\chi_{xy}|^{2}} \; .
	\end{equation}

In Fig. 5 we illustrate the different frequency dependences of the 
magnetic contrast $d$ and the absorptive part of the magnetic susceptibility 
tensor element Re$(\chi_{xy})$. 
We find a strong contrast of more than 70\% near the frequency of the abrupt 
change in $\varphi $
where the spin-polarized absorption depends only on the electronic 
structure. 
The maximum in Re$(\chi_{xy})$ occurs at $\Delta E =0.67$ eV. 
In previous more detailed microscopic calculations of $\chi_{xy}$~\cite{sursc} 
we found a linear dependence of amplitudes of pronounced maxima in $\chi_{xy}$ 
on the magnetic moment, which seems to be a general property. On 
\vspace*{-8.5cm}
\begin{figure}
\epsfxsize=10.2cm
\centerline{\quad \epsfbox{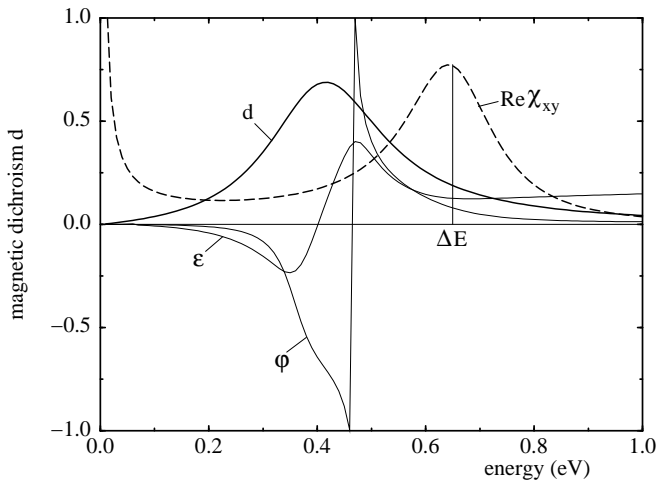}}
\vskip  1.0cm
\caption{Frequency dependence of the magnetic dichroism $d$ and the 
absorptive part of the magnetic susceptibility $Im(\chi_{xy})$.}
\end{figure}
\bigskip
\noindent
the other hand, inspecting Eq.(~\ref{lin}) (small Kerr effect) 
for a weakly varying denominator we find $\varphi \sim \chi_{xy} \sim M$. 
For the $\pm 90^{\circ }$ change in $\varphi $, by contrast, the Kerr rotation 
as well as the dichroism are determined by nonmagnetic optical properties 
not reflecting the magnetization of the material. 
Thus, the 180$^{\circ}$ `Kerr rotation' turns out to be a case of
polarization rotation in a magnetic material 
not simply related to magnetic properties. 
Experimentally, the usual MOKE and the here discussed polarization rotation 
can be distinguished by an additional analysis of the optical 
reflectivity.  

It would be of interest to determine also the {\em nonlinear} Kerr rotation 
in CeSb. 
Generally, the nonlinear Kerr effect involves more transitions and will be a 
more sensitive probe of the electronic structure. 
Also, the nonlinear Kerr rotation would illustrate the different nature of 
giant Kerr rotations in linear and nonlinear optics. 
The theoretical analysis for this follows from previous 
studies~\cite{pusto,Hue95}. 

Finally, it is of interest to point out the important formal aspect of our 
analysis, namely that Eqs. (4) and (6) for the Kerr rotation offer quite 
generally the possibility of large Kerr rotations. 
The conditions for this are discussed. 
Thereby, we analyzed the reflected electrical field with arbitrary ellipticity 
for the up to now not discussed case of Re$(\kappa ) =0$. 
We showed the importance of the intraband term in $\chi_{xx}$ and analyzed 
quantitatively its influence to the $\pm 90^{\circ }$ rotation angle. 

We thank J. Schoenes and P. Wachter 
for stimulating discussions.  

%
%

%

\begin{references}
\bibitem[9]
{einheit} All electrodynamical relations are given in SI units.
\bibitem[11]
{foot} An equivalent solution of Eq.~(\ref{glw1}) is given by 
Groot Koerkamp~\cite{Koerdipl} with
	\[
	\varphi \; = \; \frac {1} {2} \; \arctan \left( 
	\frac {2 Re(\kappa )} {1-|\kappa |^{2} } \right) \; .
	\]
The solution of Eq. (1) for $\varphi $ and $\varepsilon $ starts with 
rewriting this complex equation into two real equations for $Re(\kappa )$ 
and $Im(\kappa )$. 
Whereas Eq. (2) was found by straightforward substitution of $\tan 
\varepsilon $, 
another solution is found adding $|Re(\kappa )|^{2}$ and $|Im(\kappa )|^{2}$ 
and subsequently using the addition theorem between 
$\tan \varphi $ and $\tan 2\varphi $. 
Then, the following inversion of $\tan 2\varphi $ yields an unphysical 
halving of the definition range to $[-\pi /4,\pi /4]$, which is compensated  
by the angle $\varphi_{0} $ extending the definition range to 
$[-\pi /2 , \pi /2]$.
\bibitem[13]
{dipol} Note, in magneto-optics the observation of $\Delta l = \pm 1$ for dipole 
transitions is not required due to the addition of spin-orbit 
interaction. Furthermore, due to hybridization the discussed bands contain 
beside the $p$ character also some $f$-electron character.
\bibitem[15]
{Bemerkung} Note, these conditions change when transitions for both majority 
and minority spin bands have to be considered.
\end{references}
\end{document}